# Current Distribution Models for the Earth's Main Magnetic Field: A Discrete Inverse Theory Approach

Terence V. Sewards
Sandia Research Center
21 Perdiz Canyon Road
Placitas, N.M.  87043

tsewards@yahoo.com

**Abstract**

Current source models for the Earth's main geomagnetic field are calculated employing conventional discrete inverse theory methods. Source structures are spherical surfaces placed at the surface of the Earth's core, and at the surface of the Earth. The data set consists of measurements taken by the MAGSAT satellite in 1979.

**1. Introduction**

The Earth's main magnetic field is generated by an electric current distribution, presumably located within the Earth or near the Earth's surface. At the present time, it is generally thought that the current distribution lies in the core, and that it is generated by a magnetohydrodynamic dynamo. Determining the structure of this current distribution is an essential problem in geomagnetism. However, while it is possible to calculate the magnetic field for any given current distribution using the Biot-Savart law, the opposite problem - determining the current structure from measurements of the magnetic field - is less straightforward. To date, current source models that predict the main geomagnetic field in which the geometry of the source is prescribed initially have been limited to current loops (Peddie, 1979; Zidarov and Petrova, 1979; Alldredge, 1980; 1987). The usual procedure is to select a number of current loops and vary their position, orientation, and strength until the best possible configuration is obtained. Another method, discussed originally by Chapman and Bartels (1940), involves the use of a spherical surface current stream function $\psi(\theta,\phi)$ that can be expressed as contours on a sphere, expanding this function in a spherical harmonic series, and then directly relating the expansion coefficients to the Gaussian coefficients $g_l^m$ and $h_l^m$ (Chapman and Bartels, 1940; p.630; Stump and Polack, 1998; Lowes and Duka, 2011). The components of the current distribution, $J_\phi$ and $J_\theta$, can then be calculated by differentiation of the stream function $\psi(\theta,\phi)$ with respect to $\theta$ and $\phi$ (Stump and Pollack, 1998).

This work addresses the problem of determining models of current distributions that accurately predict the main geomagnetic field obtained from the MAGSAT satellite, employing conventional discrete inverse theory methods (Tarantola, 1987; Menke, 1989).

## 2. Inverse Theory Formulation

The geomagnetic forward problem is governed by the Biot-Savart law, through which one obtains the vector magnetic induction $\mathbf{B}(\mathbf{r}_0)$ at the field point $\mathbf{r}_0$ as a function of the vector current density $\mathbf{J}(\mathbf{r}_1)$ at the source point $\mathbf{r}_1$:

$$\mathbf{B}(\mathbf{r}) = \frac{\mu_0}{4\pi} \int_{v_1} \frac{1}{R^3} \mathbf{J}(\mathbf{r}_1) \times (\mathbf{r}_0 - \mathbf{r}_1) dv_1 \quad (1)$$

where $R = |\mathbf{r}_0 - \mathbf{r}_1|$. In Cartesian components,

$$\mathbf{J}(\mathbf{r}_1) = \mathbf{a}_x J_x + \mathbf{a}_y J_y + \mathbf{a}_z J_z \quad (2)$$

$$R = [(x_0 - x_1) + (y_0 - y_1) + (z_0 - z_1)]^{1/2} \quad (3)$$

The Cartesian coordinates of $\mathbf{B}(\mathbf{r})$ may be evaluated from Eq. 1:

$$B_x(x_0, y_0, z_0) = -\frac{\mu_0}{4\pi} \int_{v_1} \frac{1}{R^3} \left[ J_{z_1}(y_0 - y_1) - J_{y_1}(z_0 - z_1) \right] dv_1$$

$$B_y(x_0, y_0, z_0) = -\frac{\mu_0}{4\pi} \int_{v_1} \frac{1}{R^3} \left[ J_{x_1}(z_0 - z_1) - J_{z_1}(x_0 - x_1) \right] dv_1 \quad (4)$$

$$B_z(x_0, y_0, z_0) = -\frac{\mu_0}{4\pi} \int_{v_1} \frac{1}{R^3} \left[ J_{y_1}(x_0 - x_1) - J_{x_1}(y_0 - y_1) \right] dv_1$$

Since the geomagnetic inverse problem has spherical symmetry, it is useful to transform to spherical coordinates. This results in the following expressions for the field components:

$$B_r = -\frac{\mu_0}{4\pi} \int_{v_1} [k_r J_{r_1} + l_r J_{\theta_1} + m_r J_{\phi_1}] dv_1$$

$$B_\theta = -\frac{\mu_0}{4\pi} \int_{v_1} [k_\theta J_{r_1} + l_\theta J_{\theta_1} + m_\theta J_{\phi_1}] dv_1 \quad (5)$$

$$B_\phi = -\frac{\mu_0}{4\pi} \int_{v_1} [k_\phi J_{r_1} + l_\phi J_{\theta_1} + m_\phi J_{\phi_1}] dv_1$$

where

$$k_r = 0$$
$$l_r = -r_1 \sin\theta_0 \sin(\phi_0 - \phi_1)/R^3$$
$$m_r = -r_1[\sin\theta_0 \cos\theta_1 \cos(\phi_0 - \phi_1) - \cos\theta_0 \sin\theta_1]/R^3$$

$$k_\theta = r_0 \sin\theta_1 \sin(\phi_0 - \phi_1)/R^3$$
$$l_\theta = -[r_1 \cos\theta_0 - r_0 \cos\theta_1] \sin(\phi_0 - \phi_1)/R^3 \qquad (6)$$
$$m_\theta = -\{r_0 \cos(\phi_0 - \phi_1) - r_1[\cos\theta_0 \cos\theta_1 + \sin\theta_0 \sin\theta_1]\}/R^3$$

$$k_\phi = r_0[\cos\theta_0 \sin\theta_0 \cos(\phi_0 - \phi_1) - \sin\theta_0 \cos\theta_1]/R^3$$
$$l_\phi = -\{[r_1 \cos(\phi_0 - \phi_1) - r_0[\cos\theta_0 \cos\theta_1 \cos(\phi_0 - \phi_1) + \sin\theta_0 \sin\theta_1]\}/R^3$$
$$m_\phi = \{[r_0 \cos\theta_0 - r_1 \cos\theta_1]\sin(\phi_0 - \phi_1)\}/R^3$$

Equations (5) and (6) constitute the Biot-Savart law expressed in spherical coordinates. The $k_x, l_x, m_x$ coefficients were first determined by Kisabeth and Rostoker (1977). In these expressions the distance $R$ between the source and field points is given by

$$R = [r_0^2 + a^2 - 2ar_0 \cos\lambda]^{1/2} \qquad (7)$$

where

$$\cos\lambda = \cos\theta_0 \cos\theta_1 + \sin\theta_0 \sin\theta_1 \cos(\phi_0 - \phi_1) \qquad (8)$$

The fact that $k_r$ is zero means that a radial source current does not contribute to the production of the radial magnetic field component.

The geometry of the source area needs to be specified, and, for the sake of simplicity, a spherical surface is chosen, such that $J_{r_1} = 0$. The radius of this surface is then selected so that the current is restricted to (a) the surface of the core, and (b) the surface of the Earth, for the purpose of comparison. In order to obtain a continuous model current distribution in the source area while using a discrete set of parameters, the surface current vector $\mathbf{J}(\theta, \phi)$ is expanded in a spherical harmonic series

$$\mathbf{J}(\theta_1, \phi_1) = \sum_{l=0}^{5} \sum_{m=0}^{l} \{A_{lm} \mathbf{L}[P_l^m(\theta_1) \cos(m\phi_1)] + B_{lm} \mathbf{L}[P_l^m(\theta_1) \sin(m\phi_1)]\} \qquad (9)$$

This (vector spherical harmonic) expansion ensures that the divergence of the current vector is zero ($\nabla \cdot \mathbf{J} = 0$) (Stump and Pollack, 1998). The vector operator $\mathbf{L}$ is given by

$$\mathbf{L} = \mathbf{r}_1 \times \nabla = -\hat{a}_{\theta_1} \frac{1}{\sin\theta_1} \frac{\partial}{\partial \phi_1} + \hat{a}_{\phi_1} \frac{\partial}{\partial \theta_1} \tag{10}$$

In equation (9) the terms $P_l^m(\theta_1)$ are the Schmidt-normalized associated Legendre polynomials. The two components of $\mathbf{J}(\theta_1,\phi_1)$ are thus

$$J_{\theta_1}(\theta_1,\phi_1) = \sum_{l=0}^{l\max}\sum_{m=0}^{l} \frac{m}{\sin\theta_1} P_l^m(\theta_1)\left\{A_{lm}\sin m\phi_1 - B_{lm}\cos m\phi_1\right\}$$

$$J_{\phi_1}(\theta_1,\phi_1) = \sum_{l=0}^{l\max}\sum_{m=0}^{l} \frac{\partial P_l^m}{\partial \theta_1}\left\{A_{lm}\cos m\phi_1 + B_{lm}\sin m\phi_1\right\} \tag{11}$$

where *lmax* is the maximum order of the current spherical harmonic expansion, herein chosen to be 8 in order to exclude crustal contributions. The corresponding number of model parameters $N_m$ is then 88. Substitution of these expressions into equations (5), with $J_{r_1} = 0$, yields

$$B_r(\theta_0,\phi_0) = -\frac{\mu_0}{4\pi} \sum_{L=1}^{l\max}\sum_{m=0}^{l} [I_{lm}^{(1)}A_{lm} + I_{lm}^{(2)}B_{lm}] \tag{12a}$$

$$B_\theta(\theta_0,\phi_0) = -\frac{\mu_0}{4\pi} \sum_{l=1}^{l\max}\sum_{m=0}^{l} [J_{lm}^{(1)}A_{lm} + J_{lm}^{(2)}B_{lm}] \tag{12b}$$

$$B_\phi(\theta_0,\phi_0) = -\frac{\mu_0}{4\pi} \sum_{l=1}^{l\max}\sum_{m=0}^{l} [K_{lm}^{(1)}A_{lm} + K_{lm}^{(2)}B_{lm}] \tag{12c}$$

where

$$I_{lm}^{(1)} = \int_{s1} [l_r \frac{m}{\sin\theta_1} P_l^m(\theta_1)\sin m\phi_1 + m_r \frac{dP_l^m(\theta_1)}{d\theta_1}\cos m\phi_1]ds_1 \tag{13a}$$

$$I_{lm}^{(2)} = \int_{s1} [-l_r \frac{m}{\sin\theta_1} P_l^m(\theta_1) \cos m\phi_1 + m_r \frac{dP_l^m(\theta_1)}{d\theta_1} \sin m\phi_1] ds_1 \quad (13b)$$

$$J_{lm}^{(1)} = \int_{s1} [l_\theta \frac{m}{\sin\theta_1} P_l^m(\theta_1) \sin m\phi_1 + m_\theta \frac{dP_l^m(\theta_1)}{d\theta_1} \cos m\phi_1] ds_1 \quad (13c)$$

$$J_{lm}^{(2)} = \int_{s1} [-l_\theta \frac{m}{\sin\theta_1} P_l^m(\theta_1) \cos m\phi_1 + m_\theta \frac{dP_l^m(\theta_1)}{d\theta_1} \sin m\phi_1] ds_1 \quad (13d)$$

$$K_{lm}^{(1)} = \int [l_\phi \frac{m}{\sin\theta_1} P_l^m(\theta_1) \sin m\phi_1 + m_\phi \frac{\partial P_l^m(\theta_1)}{\partial \theta_1} \cos m\phi_1] ds_1 \quad (13e)$$

$$K_{lm}^{(2)} = \int [-l_\phi \frac{m}{\sin\theta_1} P_l^m(\theta_1) \cos m\phi_1 + m_\phi \frac{\partial P_l^m(\theta_1)}{\partial \theta_1} \sin m\phi_1] ds_1 \quad (13f)$$

For a data set that is restricted to values of the radial field, only equations (12a), (13a) and (13b) pertain. If it consists of $N_d$ values of $B_r(r_0, \theta_0, \phi_0)$ distributed over the surface of the spherical current source, then for each magnetic datum there are corresponding equations (12a), (13a) and 13b).

The summations over $l$ and $m$ in equations (11) can be replaced by a single summation over another index, here chosen to be $j$. That is to say,

$$\sum_l \sum_m \to \sum_j \quad (14)$$

The necessary relationship between the indices $l$, $m$ and $j$ is given by

$$j = \frac{l^2 + l + 2}{2} + m \quad (15)$$

The coefficients $A_l^m$ and $B_l^m$ can now be arranged as pairs of numbers $A_j$, $B_j$, which are then organized into a single column vector **m**:

$$\mathbf{m} = [A_1, B_1; A_2, B_2; A_3, B_3; \ldots\ldots]^T \quad (16)$$

Each of the above equations (12a, 12b, 12c) can now be expressed in matrix form:

$$\mathbf{d} = \mathbf{G}\,\mathbf{m} \qquad (17)$$

In the case of equation (12a) for the radial field, the matrix elements $G_{ij}$ of $\mathbf{G}$ are given by

$$G_{ij}^{(A)} = -\frac{\mu_0}{4\pi} I_1 \qquad G_{ij}^{(B)} = -\frac{\mu_0}{4\pi} I_2 \qquad (18)$$

Equation (17) represents the standard linear inverse problem of discrete inverse theory (Menke, 1989). Although solutions $\mathbf{m}$ of the problem yield a finite set of numbers $m_1, m_2, \ldots, m_{N_m}$, because the model is based on spherical harmonic expansions the end result is a continuous current model on the spherical surface.

### 3. Inversion

When the problem posed by equation (17) is overdetermined, the usual way to invert it is to employ either the least squares or singular value decomposition (SVD) method. Here the latter method is chosen, although it was found that least squares solutions are very similar to those obtained by the SVD method. In the SVD paradigm, the kernel matrix $\mathbf{G}$ is decomposed as follows:

$$\mathbf{G} = \mathbf{U}\,\mathbf{S}\,\mathbf{V}^T \qquad (16)$$

in which $\mathbf{U}$ is $N_d \times N_d$ and $\mathbf{V}$ is $N_m \times N_m$. The matrix $\mathbf{S}$ is $N_d \times N_m$ and diagonal, its elements are the singular values of $\mathbf{G}$. These are arranged in decreasing order, and truncated at the $p^{th}$ term in order to stabilize the inversion when the matrix $\mathbf{G}$ is ill-conditioned (Menke, 1989; Hansen, 1990). The solution is then

$$\mathbf{m}_{est} = \mathbf{V}_p\,\mathbf{S}_p^{-1}\,\mathbf{U}_p^T\,\mathbf{d} \qquad (17)$$

where $\mathbf{V}_p$ and $\mathbf{U}_p$ consist of the first $p$ columns of $\mathbf{V}$ and $\mathbf{U}$ respectively, $\mathbf{S}_p^{-1}$ is the inverse of the truncated (square) singular value matrix $\mathbf{S}_p$, and $\mathbf{m}_{est}$ is the vector of estimated model parameters.

## 4. Data Set

The data employed in this study are 806 globally distributed MAGSAT measurements of the vertical ($B_r$) field taken on Nov. 4 - 7, 1979. These were magnetically quiet days, with low values for the parameter $K_p$. They were chosen from a reduced data set of 15,837 points provided by R.A. Langel of Goddard Space Flight Center, from which anomalous points had been removed. The reduction from 15,837 to 806 points was accomplished by specifying an even spacing of approximately 2250 Km between adjacent (consecutive, in terms of spacecraft motion) points. The r.s.s. error for the $B_r$ values is approximately 6 nT (Langel and Estes, 1985). Details of the MAGSAT mission, including the spacecraft's trajectory, are described in Langel (1982) and Langel et al. (1982).

## 5. Results

As mentioned above, the two source current distributions chosen were the surface of the core, and the surface of the Earth itself. The numerical computation of the singular value decomposition was executed using the Scilab software (Scilab Enterprises, 2012).

### 5.1 Core Surface Model

The chosen singular value cutoff value for this model ($p = 80$) excludes eight near-zero values. The total prediction error for this value of $p$, which is given by the formula $E = (\mathbf{d} - \mathbf{G}\mathbf{m}_{est})^T (\mathbf{d} - \mathbf{G}\mathbf{m}_{est})$ is about 1.65 x 10$^6$ nT$^2$, so the r.m.s. prediction error is $\langle e \rangle = [E/N_d]^{1/2} \approx 45$ nT. Contours of the current densities $J_{\phi_1}$ as a function of $\theta_1$ and $\phi_1$ are shown in Fig. 1a. As may be seen in the chart, $J_{\phi_1}$ is characterized by a fairly broad band of negative current at North-equatorial latitudes, a narrower band of positive current at northern latitudes, and a mixed negative/positive band at Southern latitudes. The contour map for $J_{\theta_1}$ (Fig. 1b) is more complex, with patches of positive and negative current distributed over the surface of the core.

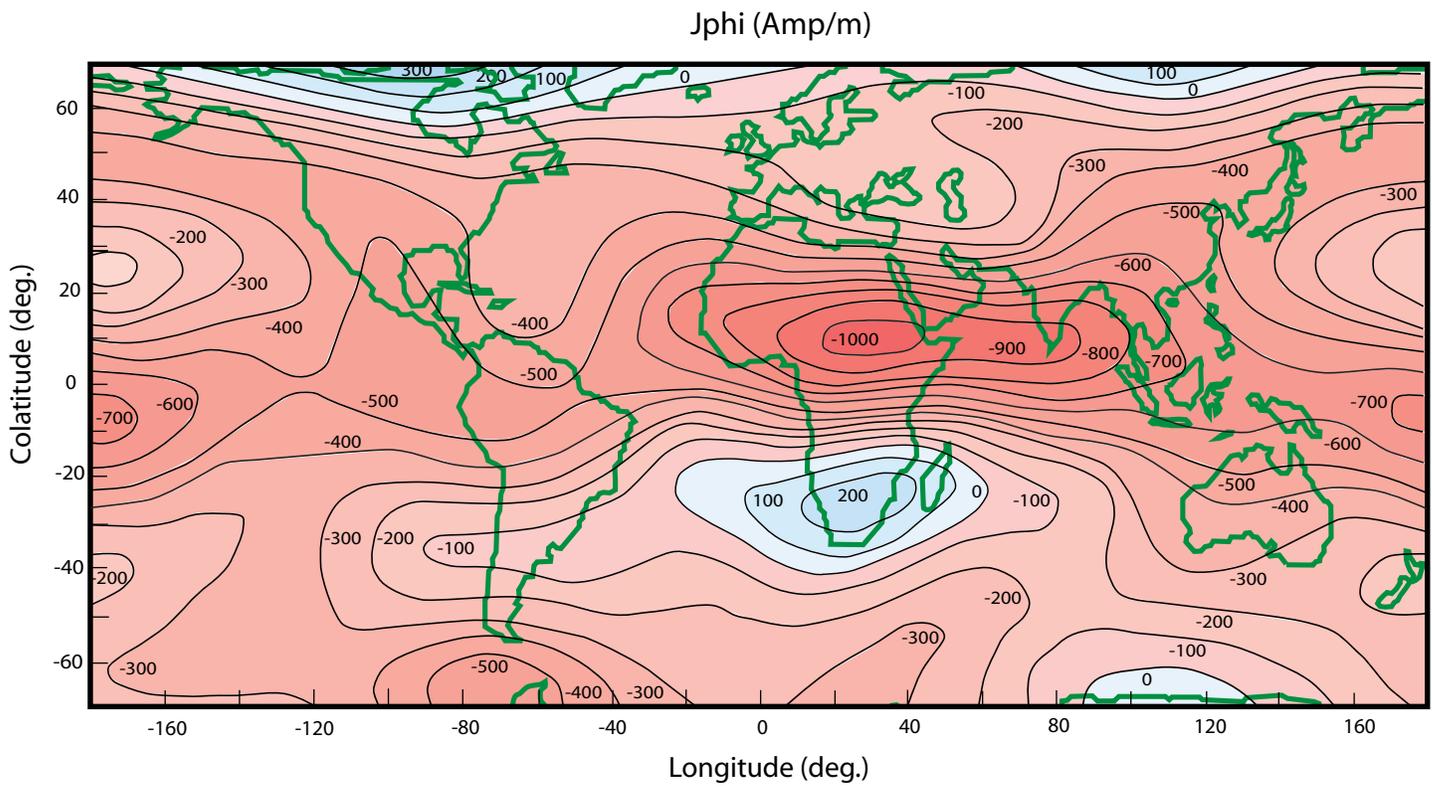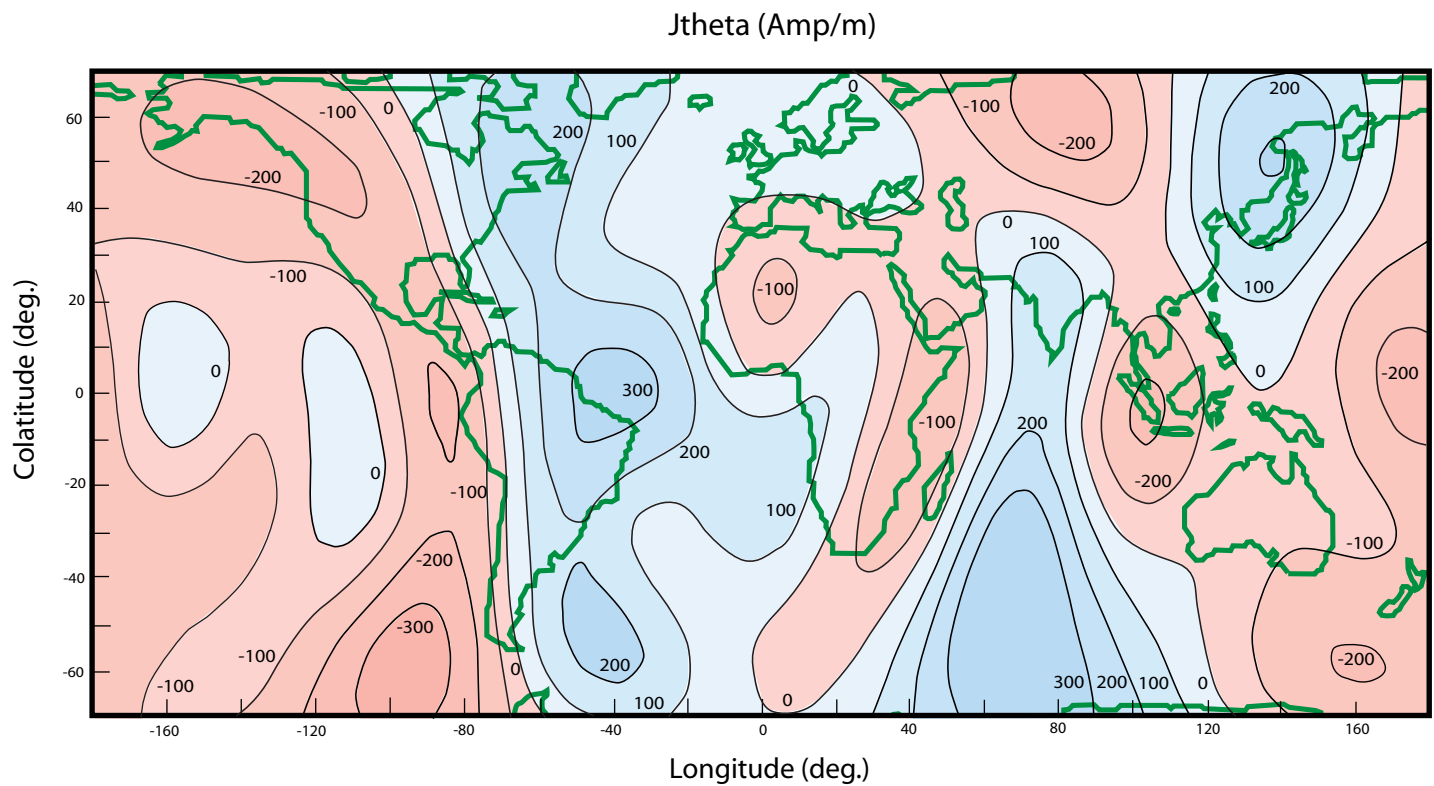

Figures 1a and 1b. Model spherical surface currents Jphi (top) and Jtheta (bottom) at the CMB which generate the geomagnetic field as measured by the MAGSAT satellite.

## 5.2 Earth Surface Model

The singular value cutoff value was again 80, which again excluded eight near-zero values. The total prediction error $E$ is 1.03 x $10^7$ nT$^2$, corresponding to an r.m.s. error of $\langle e \rangle = 71.53$ nT. Contours of the current distribution $J_{\phi_1}$ as a function of $\theta_1$ and $\phi_1$ is shown in Fig. 2a. In comparison with the corresponding distribution for the core surface model, the Earth surface $J_{\phi_1}$ distribution simpler, but roughly retains the same overall geometry. It is everywhere negative, and is concentrated at mid-latitudes, with a strong maximum over Southeast Asia. The plot is similar to the contour map of the $B_\theta$ component of the 1980 GSFC(12/83) model at the Earth's surface (Langel, 1987), which was also computed from MAGSAT data. The only significant difference is that the latter map has a region in the Antarctic where the sign of the field is reversed, whereas the $J_{\phi_1}$ map is negative everywhere. The $J_{\theta_1}$ contours are somewhat more complex (Fig. 2b), and resemble those for the $B_\phi$ component of the 1980 GSFC(12/83) model (Langel, 1987). Overall, the absolute magnitude of the $J_{\theta_1}$ current is significantly lower than the magnitude of $J_{\phi_1}$ (Fig. 2b).

## 6. Discussion

Modeling the geomagnetic field with spherical surface current distributions placed within the Earth's core and on its surface yields solutions that are much more realistic than those provided by current loop models. The method applied here is not limited to models in which the source structure is 2-dimensional and spherical.

Extension to spherical partial or full solid spheres is easily accomplished by expanding the current in the radial dimension using spherical Bessel functions. More generally, any continuous source structure that has spherical, cylindrical, or rectangular symmetry can be modeled by series expansions, and after truncation yields a finite set of parameters that effectively describes the continuous model. Then, if the data consists of a finite set of measurements, the inverse problem can be expressed in the form of equation (**14**), and a solution obtained via the SVD method as in equation (**17**), or by some other method. This his highly desirable, as the discrete inverse theory method is considerably simpler than its continuous counterpart (Menke, 1989). In this study, however, it was found that solid sphere (e.g. whole core) current models produced considerably larger errors

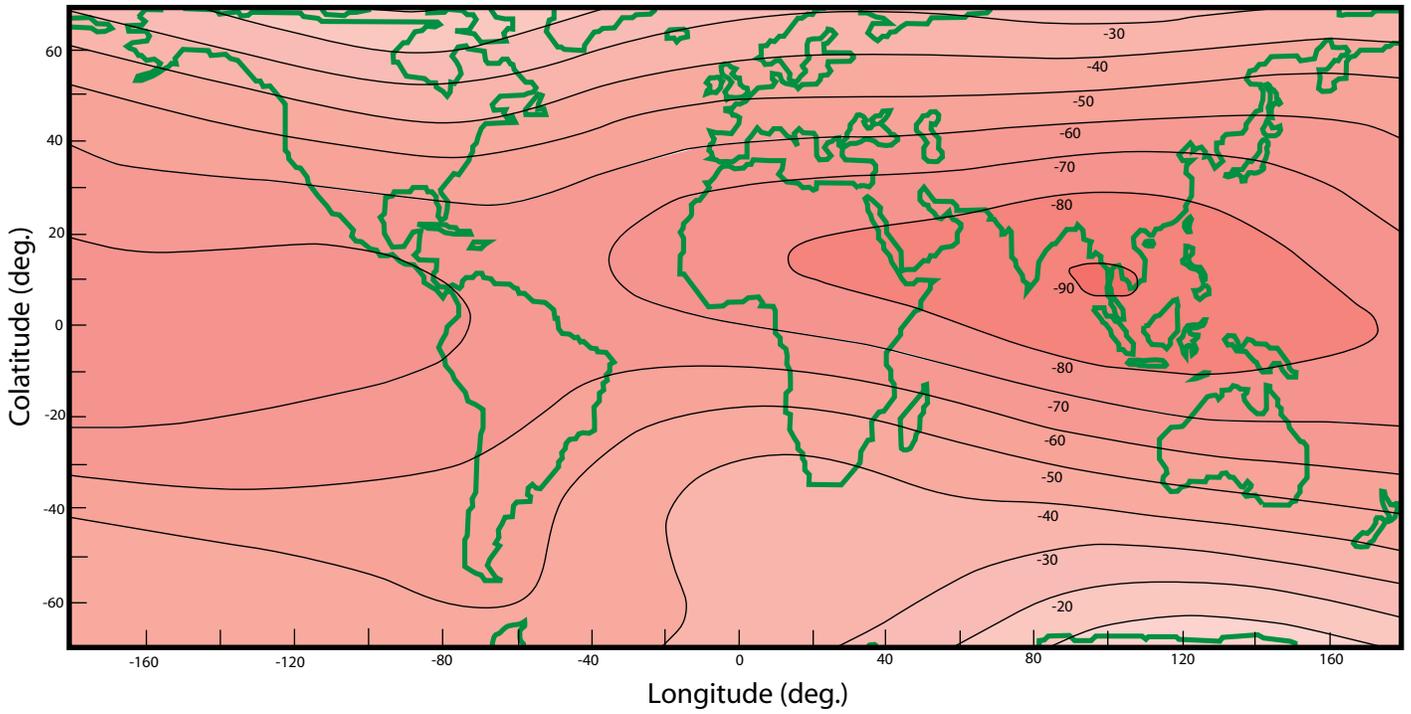

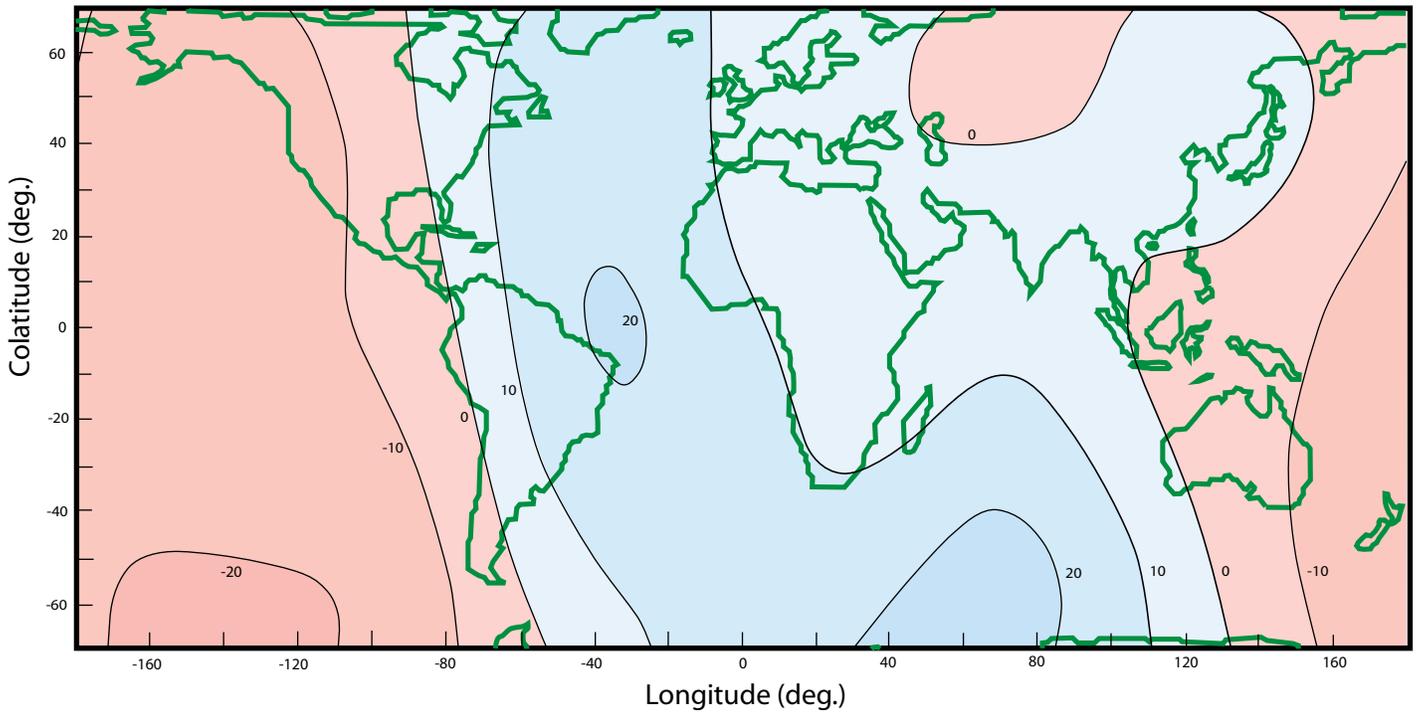

Figures 2a and 2b. Model spherical surface currents Jphi (top) and Jtheta (bottom) at the Earth's surface which generate the geomagnetic field as measured by the MAGSAT satellite.

than current surface models, and the current was concentrated at the top of the sphere.

The problem of the non-uniqueness of solutions to inverse theory problems has been discussed in some detail in the literature (Backus and Gilbert, 1970; Parker, 1977; Menke, 1989). In the case of the main geomagnetic field, if the source current lies within the core, a poloidal core current structure will produce a toroidal magnetic field which is largely or completely confined to the core, and therefore will not be measurable at the Earth's surface. Hence the currents that actually produce the magnetic field measured at the surface could be minor components of a much larger core current structure. The only limitation on the magnitude of such poloidal currents is due to the ohmic heating constraint. Evidently the surface current models obtained here, which are based on the "natural" (SVD) solution to the geomagnetic inverse problem, are in essence the minimal or simplest current structures that will produce the measured field.